\documentclass[structabstract]{aa}
\usepackage{natbib}			% BibTeX style bibliography
\usepackage{amsmath}			% Mathematic formulaes
\usepackage{graphicx} 			% Pictures and images
\usepackage[varg]{txfonts}		% A&A is printed using the Postscript TX Times-fonts.
\usepackage[english]{babel}		% Language is english
\usepackage{nameref}			% References
\usepackage[normalem]{ulem} 		% for striketrough type effects
\usepackage{paralist}			% lists

\begin{document}
  \title{Observational evidence for two distinct giant planet populations}

  \author{N. C. Santos\inst{1,2}
	  \and V. Adibekyan\inst{1}
  	  \and P. Figueira\inst{1}
	  \and D. T. Andreasen\inst{1,2}
	  \and S. C. C. Barros\inst{1}
	  \and E. Delgado-Mena\inst{1}
	  \and O. Demangeon\inst{1}
	  \and J. P. Faria\inst{1,2}
	  \and M. Oshagh\inst{3}
	  \and S. G. Sousa\inst{1}
	  \and P. T. P. Viana\inst{1,2}
	  \and A.C.S. Ferreira\inst{1,2}
	 }

  \institute{
  	  Instituto de Astrof\'isica e Ci\^encias do Espa\c{c}o, Universidade do Porto, CAUP, Rua das Estrelas, 4150-762 Porto, Portugal
	  \and
	  Departamento de F\'isica e Astronomia, Faculdade de Ci\^encias, Universidade do Porto, Rua do Campo Alegre, 4169-007 Porto, Portugal
	  \and
	  Institut f\"ur Astrophysik, Georg-August-Universit\"at, Friedrich-Hund-Platz 1, 37077 G\"ottingen, Germany
}

  \date{Received date / Accepted date }
%----------------------------------------------------------------------------------------
%	Abstract
%----------------------------------------------------------------------------------------
  \abstract
  {Analysis of the statistical properties of exoplanets, together with those of their host stars, are providing a unique view into the process of planet formation and evolution.}
  {In this paper we explore the properties of the mass distribution of giant planet companions to solar-type stars, in a quest for clues about their formation process.} 
  {With this goal in mind we studied, with the help of standard statistical tests, the mass distribution of giant planets using data from the exoplanet.eu catalog and the SWEET-Cat database of stellar parameters for stars with planets.}
  {We show that the mass distribution of giant planet companions is likely to present more than one population with a change in regime around 4\,M$_{Jup}$. Above this value host stars tend to be more metal poor and more massive and have [Fe/H] distributions that are statistically similar to those observed in field stars of similar mass. On the other hand, stars that host planets below this limit show the well-known metallicity-giant planet frequency correlation.}
  {We discuss these results in light of various planet formation models and explore the implications they may have on our understanding of the formation of giant planets. In particular, we discuss the possibility that the existence of two separate populations of giant planets indicates that two different processes of formation are at play. }

%----------------------------------------------------------------------------------
%	Title
%----------------------------------------------------------------------------------

  \maketitle
%---------------------------------
  
  -------------------------------------------------
%	Introduction
%----------------------------------------------------------------------------------
  \section{Introduction}					\label{sec:Introduction}

The detection of more than 3500 planets orbiting solar-type stars \citep[e.g. exoplanet.eu --][]{Schneider-2011} makes the exoplanet domain prone to statistical studies. This opens the possibility to analyse the properties of newly found worlds and, in comparison with the model predictions, better understand the processes of planet formation and evolution \citep[see e.g.][]{Mayor-2014}. 

In this context, a significant amount of information was brought by the analysis of the star-planet connection. Initial findings have shown that the planet occurrence is closely linked to the metallicity of the host star \citep[][]{Gonzalez-1997,Santos-2001,Santos-2004b,Ida-2004b,Fischer-2005,Mordasini-2012}, even if this trend is a matter of debate for planets hosted by giant stars \citep[][]{Pasquini-2007,Reffert-2015}. Stellar mass has also been suggested to influence severely the planet formation efficiency \citep[][]{Johnson-2007,Lovis-2007,Bonfils-2013,Kennedy-2008,Reffert-2015}. Furthermore, several studies have shown that the architecture of the planetary systems is closely connected to the properties of the star. For example, it is now known that the orbital period and eccentricity of the planets may depend on the chemical content of the star \citep[][]{Dawson-2013,Beauge-2013,Adibekyan-2013}. In brief, the understanding of planet formation and evolution processes is tightly connected to the understanding of the properties of their host stars.

In this paper we explore the mass distribution of giant planets orbiting solar-type stars in a search for clues about the formation of giant planets. In Sect.\,\ref{sec:mass} we present evidence that this distribution likely presents more than one population, with two regimes separated in planet mass. In Sect.\,\ref{sec:correl} we then explore the properties of the host stars of two planet regimes, to show that a significant difference exists concerning the stellar metallicity and stellar mass. In Sect.\,\ref{sec:discussion} the results are discussed in light of the different models of planet formation and evolution.

%
% Section 2
%

  \section{The mass distribution of giant planets}				\label{sec:mass}

\begin{figure}
\begin{center}
\includegraphics[angle=0,width=1\linewidth]{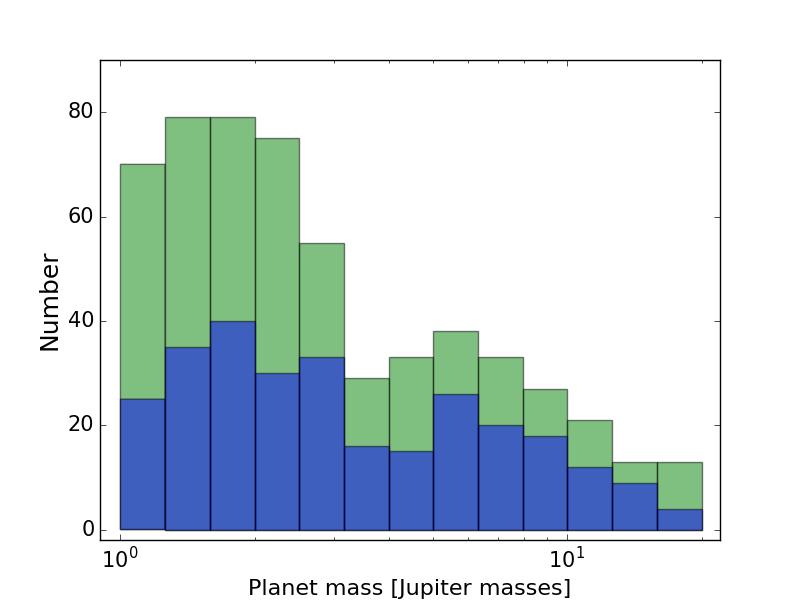}
\includegraphics[angle=0,width=1\linewidth]{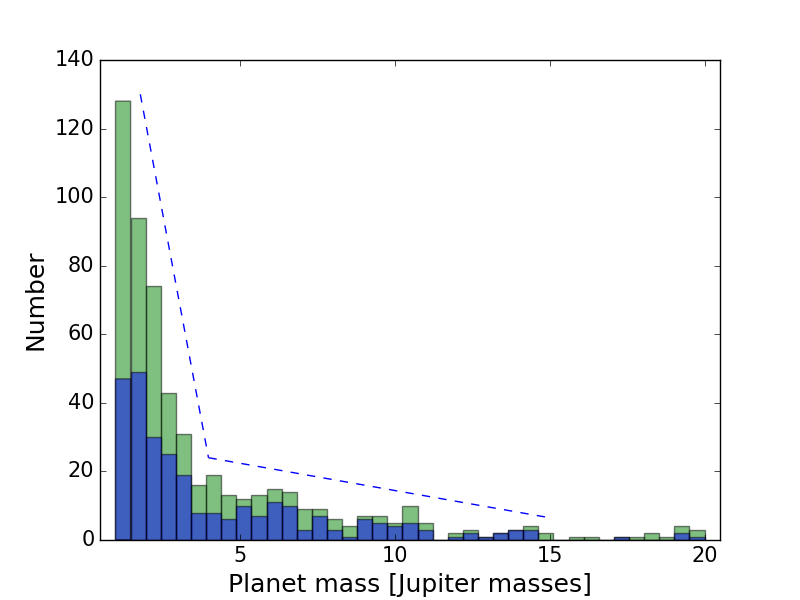}
\end{center}
\vspace{-0.5cm}
\caption{Mass distribution of giant planets around solar-type stars, in log (top panel) and linear (lower panel) scales. The green histograms include all planets in exoplanet.eu, and the blue histogram includes only planets following the criteria mentioned in Section.\,\ref{sec:correl}. }
\label{fig:mass}
\end{figure}

In Fig.\,\ref{fig:mass} we plot the mass distribution for all giant {planets with masses} between 1 and 20\,M$_{Jup}$ orbiting solar-type stars as listed in the exoplanet.eu database \citep[][]{Schneider-2011}\footnote{Data from the 16th of February 2017.}. The lower limit was set because we only want to explore the giant planet domain. We only selected planets discovered through the transit and radial velocity (RV) methods\footnote{These constitute the vast majority of the known planets with a mass estimate. {In the case of planets only detected with radial velocities, the masses represent minimum masses.}}.

In the upper panel of Fig.\,\ref{fig:mass} we plot the mass distribution in log scale, while in the lower panel we present the same distribution in linear scale. The upper panel suggests that the mass distribution may have two different maxima, separated by a valley at around 4\,M$_{Jup}$. These two regimes are also denoted in the linear scale plot: below $\sim$4\,M$_{Jup}$ the distribution has a clear peaked shape, while above this mass the distribution is suddenly mostly flat, with a slowly decreasing trend up to (tentatively) $\sim$15\,M$_{Jup}$. The two dashed lines in the plot denote this change in slope, and are used here merely as visual guide. The existence of a transition at a similar mass value ($\sim$5\,M$_{Jup}$) has also been recently mentioned by \citet[][]{Bashi-2017} based on the analysis of planets with measured mass and radius, even if these authors concentrated their discussion exclusively on planets with masses lower than this limit.

The possible existence of two mass regimes suggests that there may be two different populations of giant planets, with masses above and below $\sim$4\,M$_{Jup}$.

%
% Section 3
%

\section{Exploring the various mass regimes }				\label{sec:correl}

To explore this possibility, we compare the metallicity and mass distributions for the stars having planets in these two planet-mass regimes. For that purpose we started by compiling stellar parameters for all the stars selected above from the SWEET-Cat database\footnote{https://www.astro.up.pt/resources/sweet-cat} dabatase \citep[][]{Santos-2013}, which is a large catalog of stellar parameters for stars with planets. We then selected stars using the following criteria. First, we selected only stars for which stellar temperatures, $\log{g}$, metallicities, and masses are listed in SWEET-Cat and that have visual magnitudes lower than 13. This magnitude limit, though not very constraining, ensures that the planet masses can be derived with a reasonable confidence using radial velocities, and that the stellar parameters, namely effective temperature and metallicity, could be derived with a high level of reliability based on high quality spectroscopic data. Then, we selected only stars with temperatures below 6500\,K and above 4000\,K, thus excluding targets for which stellar atmospheric parameters could be less reliable. We also considered only planets with masses below 15\,M$_{Jup}$, to try to avoid the inclusion of brown-dwarf companions. Finally, we decided to cut in orbital period to select only planets with periods longer than 10 days and shorter than 5 years. The first limit allows us to avoid hot Jupiters, whose formation and migration processes are largely debated {\citep[e.g.][]{Ngo-2016,Nelson-2017}}. 
%{\bf Setting this limit implies that we remove from our sample most transiting planets, since transit surveys are strongly sensitive to the orbital period.} 
The upper limit allows us to guarantee that the sample is reasonably complete; giant planets in longer period orbits could have been missed in radial velocity surveys. However, we tested the results without including any constraints on the shorter period limit and adding a stronger constraint on the upper value (e.g. up to 2.5 years)\footnote{A 2\,M$_{Jup}$ planet in a five-year period circular orbit around a solar-mass star induces a semi-amplitude signal with 11\,m\,s$^{-1}$, which is a value that is within the detection capabilities of present day instrumentation; this value, however, it is not straightforward to detect if the planet is orbiting noisier, giant stars.}. The conclusions of the analysis presented below become even stronger when we include hot Jupiters or if we only select planets up to periods of just 2.5 years. A table with the selected sample is available in electronic form.

\subsection{Planet mass, stellar mass, and stellar metallicity}

\begin{figure*}[t!]
\begin{center}
\includegraphics[angle=0,width=0.4\linewidth]{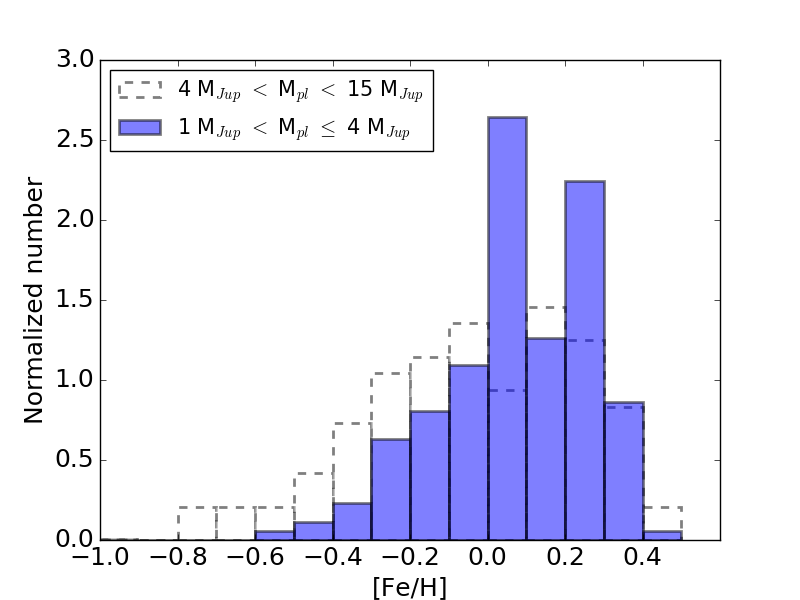}
\includegraphics[angle=0,width=0.4\linewidth]{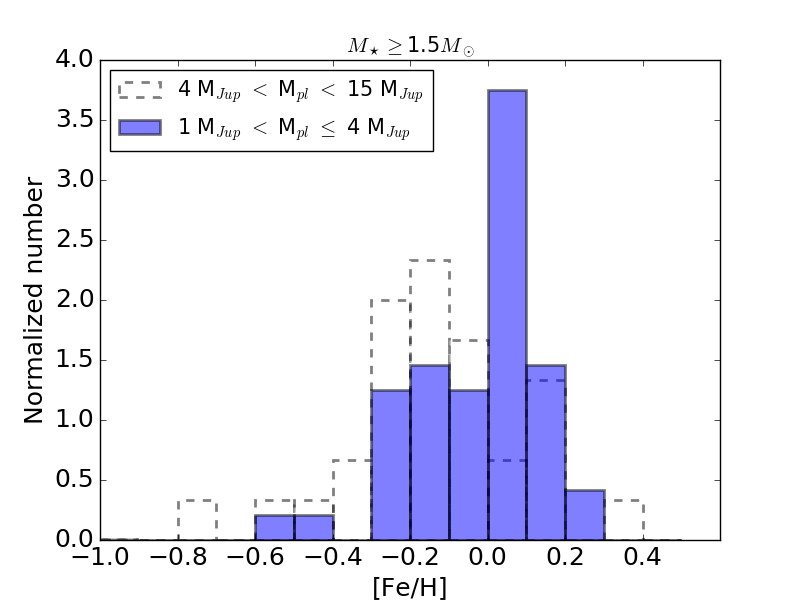}
\includegraphics[angle=0,width=0.4\linewidth]{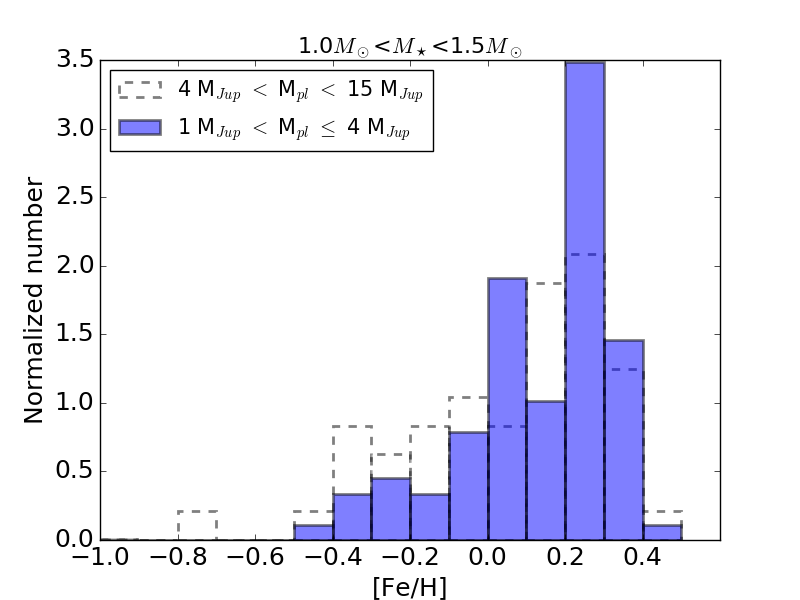}
\includegraphics[angle=0,width=0.4\linewidth]{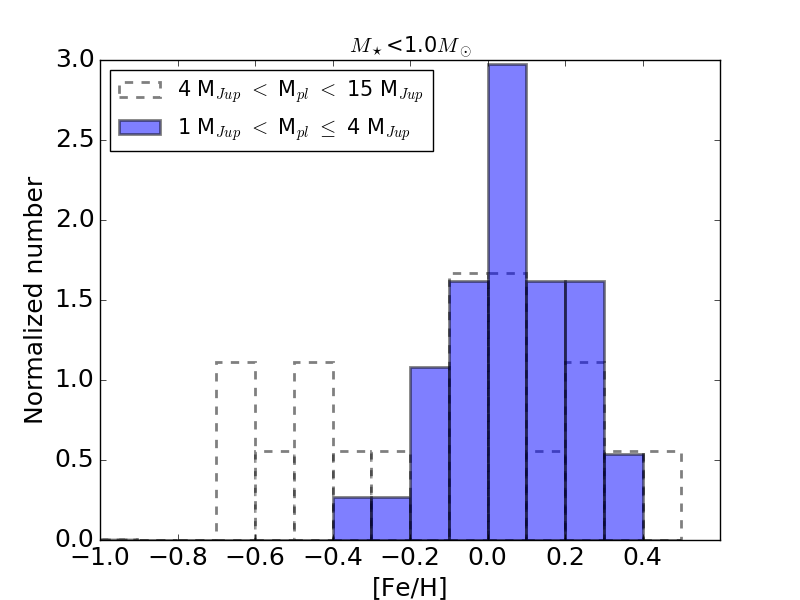}
\end{center}
\vspace{-0.5cm}
\caption{Metallicity distribution for stars with giant planets in different mass regimes. Upper left: The metallicity distribution is shown for all stars. Remaining three panels: The metallicity distribution is shown for three different stellar mass regimes.}
\label{fig:feh1}
\end{figure*}

Using this selected sample, in Fig.\,\ref{fig:feh1} we compare the [Fe/H] distribution of the stars with planet masses above and below 4\,M$_{Jup}$. {In the plots we use a ``normalized number'' to better compare the various populations (each with a different number of stars), meaning that the histograms are normalized such that the integral over the [Fe/H] range is 1.}
%Error bars in each bin were computed using the ``binomial formulation'' presented in \citet[][]{Cameron-2011}.} 
The upper left panel presents this distribution for all the hosts, while the remaining panels present the distributions for stars within different mass intervals. Stellar masses were taken from SWEET-Cat, and were derived using the calibration in \citet[][]{Torres-2010}\footnote{The obtained results do not change if we use the masses listed in exoplanet.eu.}.

For each case, we applied a Kolmogorov-Smirnov (K-S) test to explore whether the two samples (more and less massive planets) come from the same parent population\footnote{Using the python \texttt{scipy.stats.ks\_2samp} library. Similar results are obtained if we use a one-sided K-S test (\texttt{scipy.mstats.ks\_2samp}) or a Anderson-Darling test (\texttt{scipy.stats.anderson\_ksamp}).}. The results of this analysis are presented in Table\,\ref{table}, and show that if we use all the stars or only the stars with mass above 1.5\,M$_\odot$ the low p-values strongly suggest the two samples are not part of the same population. On the other hand, as we go towards lower stellar mass regimes, the K-S p-values increase, i.e., there a higher evidence for one single population \citep[see also][]{Adibekyan-2013}. However, in all stellar mass regimes the hosts of the more massive planets always have lower average metallicity values. Also, the spread of the metallicity distribution of the stars with the most massive planets, as measured via the standard deviation (STD), is always larger than that found for the lower mass planet host stars. In other words, they span a larger metallicity range.

\begin{table*}[t]
\caption{Comparison of the metallicity distributions of the stars with planets in the two mass regimes and values for the solar neighbourhood populations (for dwarfs/giants). }
\begin{center}
\begin{tabular}{c|ccc|ccc|c||ccc}
\hline
              & \multicolumn{3}{c}{$M_{pl}>4\,M_{Jup}$} &   \multicolumn{3}{c}{$M_{pl}<4\,M_{Jup}$} & & \multicolumn{3}{c}{Solar neighbourhood}\\
\hline
Sample                              & N & <[Fe/H]>  & STD & N & <[Fe/H]>  & STD & K-S p-value & N & <[Fe/H]>  & STD\\
\hline
All stars &  96  &    $-$0.04              & 0.28 &  174 & 0.07 & 0.19 & 5\,10$^{-4}$ & & $-0.11$/$-0.08$& $0.24$/$0.18$\\
$M_\star\geq$1.5\,M$_\odot$ & 30   & $-$0.14 & 0.22 & 48 & $-$0.02 & 0.16& 3\,10$^{-3}$ & 241 & $-$0.08 & 0.16 \\
1.0\,M$_\odot$<$M_\star$<1.5\,M$_\odot$ &  48  & 0.04 &  0.26 & 89 & 0.13 & 0.20 & 0.06 & 177 & 0.04 & 0.17\\
$M_\star\leq$1.0\,M$_\odot$ &  18  & $-$0.10 & 0.34 & 37 &  0.06 & 0.15 & 0.06 & 397 & $-$0.16 & 0.24\\
\hline
\end{tabular}
\end{center}
\label{table}
\end{table*}%

\begin{figure}
\begin{center}
\includegraphics[angle=0,width=1\linewidth]{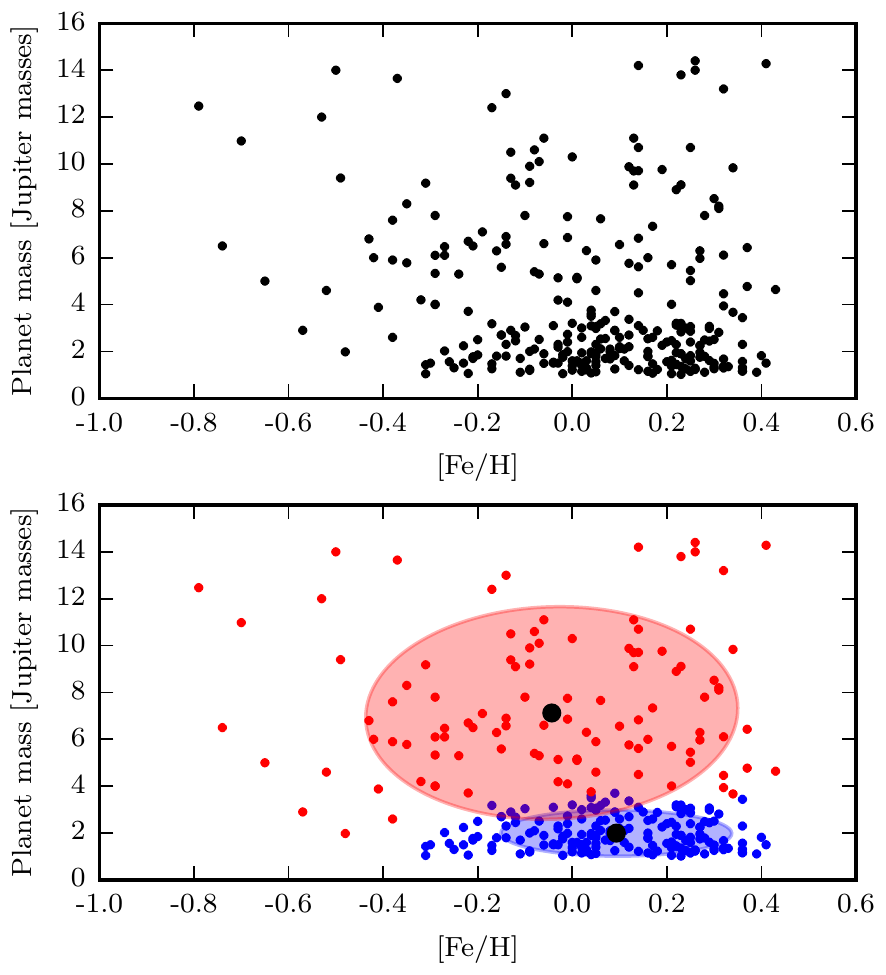}
\end{center}
\vspace{-0.5cm}
\caption{Planet mass vs. stellar metallicity plot for our sample stars. In the bottom panel we show the position of the two populations that result from our Gaussian mixture analysis (see text for more details).}
\label{fig:mix}
\end{figure}

We also ran a clustering analysis on the two-dimensional planet mass - stellar metallicity distribution using the planets selected following the criteria mentioned in the next section. The upper panel of Fig.\ref{fig:mix} shows a scatter plot of the above-mentioned variables. The plot shows a clustering for planet masses below $\sim$4\,$M_{Jup}$ and metallicities above $\sim$$-$0.3\,dex. Planets with masses above that value are more scattered in both planet mass and stellar metallicity. In the lower panel we show the result of a clustering analysis. A Gaussian mixture was considered\footnote{As implemented in the \texttt{scikit-learn} package \citep[][]{Pedregosa-2011}.}, in which the data set is modelled with a fixed number of (two-dimensional) Gaussian distributions. We assumed the presence of two clusters and iteratively optimized the parameters of the distributions using the expectation-maximization algorithm \citep[][]{Gupta-2011}. In the lower panel of Fig.\ref{fig:mix} we show the underlying distribution of our sample and the two clusters that result from this analysis. The clusters basically divide the data set into lower and higher mass planets and overlap near 4\,$M_{Jup}$. Their centres show an offset in metallicity of 0.14\,dex, where the average [Fe/H] is lower for the hosts of the more massive planets.

It is worth noting that the higher mass stars in our sample are also, on average, more evolved. The evolved stars (defined here as those with $\log{g}$<3.5) in our sample have mass values from $\sim$1 to 4.5\,M$_\odot$, while the dwarfs have maximum masses of about 1.5\,M$_\odot$ with most stars in the range between 0.7 and 1.3\,M$_\odot$. Also, above 3\,M$_\odot$ all eight stars with planets in the mass range above 4\,M$_{Jup}$ have metallicities below solar ([Fe/H] between -0.13 and -0.74\,dex), while the only star with a lower mass planet has super-solar metallicities (0.07\,dex). 

\subsection{Comparing with the solar neighbourhood}

It is interesting to see that for the most massive stars and for those with mass below solar mass, the average metallicity of the massive planet hosts is similar to the average metallicity of the solar neighbourhood for dwarfs ($-$0.11\,dex, STD=0.24) and giants ($-$0.08\,dex, STD=0.18) \citep[][]{Sousa-2011,Alves-2015}.
In other words, in these stellar mass regimes, the metallicity distribution of the stars with the most massive planets is very similar to that observed in solar neighbourhood stars. Stars in the intermediate mass range, on the other hand, present higher metallicity values, on average. This likely reflects the fact that most of these are dwarfs; as shown in \citet[e.g.][]{Santos-2013}, when a cut in colour is applied to a sample of dwarfs, more massive stars are on average more metal rich.

To explore these points in more detail, we compare the metallicity distributions of the hosts of massive planets with those of field dwarfs and giants, dividing the stars in the three groups of stellar mass as listed in Table\,\ref{table}. The comparison of the [Fe/H] distribution of the stars with $M_\star\leq$1.0\,M$_\odot$ shows that the two distributions have a K-S p-value of 0.04. Field dwarfs \citep[from][]{Sousa-2011} have slightly lower average metallicity and standard deviation. Considering dwarfs with masses between 1.0 and 1.5\,M$_\odot$, the two groups are most likely statistically similar (K-S p-value of 0.06), with the same average value ($0.04$ in both cases) but a lower dispersion for the field dwarfs. Finally, comparing the hosts of massive planets with mass above 1.5\,M$_\odot$ with giant stars of similar mass from \citet[][]{Alves-2015}, we also do not find strong hints to refute the hypothesis that the two groups may come from the same parent distribution (K-S p-value of 0.03). In this case, the average value of [Fe/H] distribution of the field stars has higher values than the one found for the host stars of the more massive planets. Again, the standard deviation has lower values.

We did the same comparison for the hosts of the lower mass planets (M$_{pl}$<4\,M$_{Jup}$). In that case, as expected, planet hosts are always strongly significantly more metal rich than field stars, in all stellar mass regimes compared. The K-S p-values are very low, ranging from 3\,10$^{-3}$ for the most massive stars, down to 10$^{-7}$ and 10$^{-8}$ for the other stellar mass regimes. The metallicity distribution of the low-mass planet hosts thus follows the usual metallicity-giant planet frequency correlation \citep[][]{Santos-2004b,Fischer-2005}.

%\begin{table*}[t]
%\caption{Comparison of the metallicity distributions of the stars with planets in the two mass regimes.}
%\begin{center}
%\begin{tabular}{c|ccc|ccc|l}
%\hline
 %             & \multicolumn{3}{c}{$M_{pl}>4\,M_{Jup}$} &   \multicolumn{3}{c}{$M_{pl}<4\,M_{Jup}$}\\
%\hline
%Sample                              & N & <[Fe/H]>  & STD & N & <[Fe/H]>  & STD & P(K-S)\\
%\hline
%All stars &  98  &    -0.04              & 0.28 &  178 & 0.08 & 0.19& 5\,10$^{-4}$\\
%$M_\star\geq$1.5\,M$_\odot$ & 30   & -0.14 & 0.22 & 50 & -0.01 & 0.16& 2\,10$^{-4}$\\
%1.0\,M$_\odot$<$M_\star$<1.5\,M$_\odot$ &  48  & 0.04 &  0.26 & 91 & 0.14 & 0.19 & 0.02 \\
%$M_\star\leq$1.0\,M$_\odot$ &  20  & -0.07 & 0.33 & 37 &  0.06 & 0.15 & 0.11\\
%\hline
%\end{tabular}
%\end{center}
%\label{table}
%\end{table*}%

%MOSTRAR GRAFICO PARA AS EVOLVED VS. NON-EVOLVED?

%\begin{figure}
%\begin{center}
%\includegraphics[angle=0,width=0.84\linewidth]{fig_mass1.png}
%\includegraphics[angle=0,width=0.84\linewidth]{fig_mass2.png}
%\end{center}
%\vspace{-0.5cm}
%\caption{Mass distribution for stars with giant planets in different mass regimes. Upper panel: for all stars. Lower panel: only for evolved stars.}
%\label{fig:mass1}
%\end{figure}

\subsection{Other planet properties}

We further tested whether any other planet property was significantly different between the two populations of planets (more and less massive than 4\,$M_{Jup}$). For instance, the orbital periods of the higher mass planet population are indeed longer: 613 versus 566 days if all stars are considered. These differences are, however, not statistically significant (K-S p-value of 0.33). It is worth mentioning, however, that as already pointed out \citep[e.g.][]{Adibekyan-2013}, lower metallicity stars host, on average, planets in longer orbital periods. As seen above, the lower-mass planet population also has higher metallicity, on average, a fact that could by itself explain this offset. 

Finally, we also tested if the eccentricity distributions of the two samples are different. The results show that no statistically significant differences exist (K-S p-value of 0.25), even if higher mass planets tend to have slightly higher average values for the orbital eccentricity when compared with the lower mass planets in our selection: 0.29 (STD=0.22) and 0.25 (STD=0.14), respectively. This trend may actually be expected from theoretical models of planet-disk interaction \citep[][]{Bitsch-2013}, as already discussed in \citet[][]{Adibekyan-2013}.

%
% Section 4
%

\section{Discussion}	  \label{sec:discussion} 

The results presented above suggest that giant planets with masses above and below $\sim$4\,M$_{Jup}$ may represent two different populations. The data we analyse shows that stars with planets more massive than 4\,M$_{Jup}$ orbit stars that are on average more metal-poor. This trend is statistically significant for the more massive stars ($M>1.5M_\odot$), even it if is also observed in all the analysed stellar mass regimes. We also show that planets with M$_{pl}$>4\,M$_{Jup}$ orbit stars that span a wider metallicity range than stars with lower mass planets, and have [Fe/H] distributions more similar to the average field stars of similar mass.

The fact that the metallicity distribution for the stars with the more massive planets is lower is intriguing. Metallicity is known to be intimately related to the frequency of giant planets, a result coming from both observations and theoretical models based on the core-accretion paradigm \citep[e.g.][]{Santos-2004b,Fischer-2005,Mordasini-2012}. More metal-rich stars may also be able to form higher mass planets, even if this trend is not necessarily strong \citep[see][and their Figs. 4 and 5]{Mordasini-2012}. The results presented here thus seem to contradict these expectations.

In the context of the core-accretion paradigm, \citet[][]{Kennedy-2008} suggested that the frequency of planets is an increasing function of stellar mass, at least up to $\sim$3\,M$_\odot$. This result is supported by observational evidence \citep[e.g.][and references within]{Reffert-2015}. Higher mass stars are also known to have higher mass disks \citep[][]{Natta-2000} that are likely capable of forming higher mass giant planets. It could thus be that the tendency for the most massive stars in our sample to host more massive planets may be explained by the simple fact that their massive disks were able to form planets with higher masses, even if their metal content was lower. This could explain, in the context of the core-accretion paradigm, the existence of two populations of planets. 

If real, however, this model would have to explain the reason for the proposed change in regime around $\sim$4\,M$_{Jup}$. Furthermore, it would have to explain why in all stellar mass regimes studied, the metallicity distributions of the hosts of planets with mass $>4\,M_{Jup}$ are always compatible with the field dwarf distribution (even if the [Fe/H] spread is always higher in the planet hosts), opposite to what is observed if we only consider hosts of the lower mass planets. For stars with mass below 1\,$M_{\odot}$, we find that planet hosts are more metal rich than field stars (even if the K-S p-value shows a value that is marginally significant). However, no difference, or even the inverse trend, is found when considering stars with mass above this limit. 

To further explore this, we compared the index $M_\star\,10^{[Fe/H]}$ of the stars with higher and lower mass planets. Assuming that the $M_\star$ is correlated with the mass of the disk, this index is expected to measure the total amount of heavy material in a disk and thus be more directly related to the planet formation efficiency in the core-accretion scenario. Comparing the two indices for higher and lower mass planets and using all the stars in our sample we conclude that planets with mass above 4\,M$_{Jup}$ have systematically lower values. A K-S test provides a p-value of 0.08. This result suggests that the increase in stellar mass only partially compensates for the difference in metallicity. However, the exact relation between disk mass and stellar mass is not fully clear \citep[e.g.][]{Andrews-2013}, and may have an impact on these results.

An alternative explanation for the observed populations calls for the disk instability process \citep[e.g.][]{Boss-1997}. It has been shown that planets formed by disk instability should in principle have higher masses and be easier to form around stars with more massive disks \citep[e.g.][]{Rafikov-2005,Nayakshin-2017}. Furthermore, it has been suggested that gravitational instability may be more efficient in more metal-poor disks (or at least not so metallicity dependent as the core-accretion process), forming planets in longer period orbits, and potentially acting as an accelerator of the core formation \citep[][]{Boss-2002,Cai-2006}.

In this scenario, the observational result presented above could be interpreted as showing the existence of two separate populations of giant planets formed by different physical processes. On the one hand, the lower mass planets (here defined with a tentative mass below 4\,M$_{Jup}$) are formed by the core-accretion process and are more prevalent around more metal-rich stars. On the other hand, the more massive planets, whose formation process is mainly done through a gravitational instability process or a process where disk instability has played a role \citep[][]{Cai-2006}. This second population is less sensitive to the stellar metallicity. {If confirmed, this scenario would imply that above $\sim$4\,M$_{Jup}$ the formation of giant planets is no longer dominated by the core-accretion process.} An overlap of the two populations likely exists, however. A deeper analysis with an increase in the number of planets in the samples is needed to confirm or refine this value.

{
In this context it is relevant to add that studies of stars with brown-dwarf companions have shown that these have metallicity distributions that are very similar to the solar neighbourhood stars \citep[][]{MataSanchez-2014,Maldonado-2017}. The existence of a so-called ``brown-dwarf desert'', a pronounced scarcity of companions around Sun-like stars with mass around $\sim$30-50\,M\,$_{Jup}$ \citep[e.g.][]{Sahlmann-2011,Ma-2014}, strongly suggests, however, that the higher mass planets are not likely to be the low-mass end of the ``stellar'' distribution.
Furthermore, recent results \citep[][]{Maldonado-2017} also suggest that a different metallicity distribution may exist for objects above and below the ``brown-dwarf desert''. This was proposed to reflect the existence of two different populations that were likely formed by different processes: disk instability for the lower mass brown dwarfs and cloud fragmentation (as ``normal'' stars) for the higher mass brown dwarfs \citep[see][]{Ma-2014}.
}

Although the metallicity-giant planet frequency correlation is a well-established fact when dealing with dwarf stars, the existence of such a correlation is still a matter of debate for giant stars. Indeed, some results suggest that this correlation may even not be present or could be weaker than that found for the dwarfs \citep[see discussions in][]{Pasquini-2007,Mortier-2013c,Maldonado-2013,Reffert-2015}. As we have also seen, the most massive stars in our sample (>1.5\,M$_\odot$) are also evolved. In the scenario mentioned above, the existence of two populations of giant planets that is observed for the higher mass stars may be deeply related to the possibility that evolved stars (more massive on average) also do not show a very clear metallicity-giant planet correlation.

A word of caution to say that this result is based on the assumption that all planets in the sample are {\it bona fide}. This may not be always the case, especially for planets orbiting giant stars \citep[see discussion in][]{Reffert-2015}.

% Spin-offs: abundancias, e feh e mass distrubition function

If these results are confirmed as new planets are detected, the discussion presented here may give a new important insight into the giant planet formation process. They also show that the study of giant planets is still of great importance, even if the focus of exoplanet research is moving towards the study of their low-mass counterparts. Giant planets discovered with the GAIA mission, whose sensitivity will allow the detection of thousands of giant planets in long period orbits around stars of different mass\citep[][]{Sozzetti-2001} may shed significant light into this case. The study of the mass-radius relation of giant planets as carried out with missions like CHEOPS \citep[][]{Fortier-2014}, TESS \citep[][]{Ricker-2010},  or PLATO \citep[][]{Rauer-2014}, as well as of their atmospheric composition using new ground- and space-based instruments such as JWST and HIRES@E-ELT \citep[e.g.][]{Greene-2016,Marconi-2016}, may also bring new constraints on the processes of planet formation \citep[][]{Fortney-2008,Mordasini-2012b,Mordasini-2016}.

%----------------------------------------------------------------------------------------
%	AcknowledgementsPapersAndBooks
%----------------------------------------------------------------------------------------
\begin{acknowledgements}

This work was supported by Funda\c{c}\~ao para a Ci\^encia e a Tecnologia (FCT, Portugal) through the research grant through national funds
and by FEDER through COMPETE2020 by grants UID/FIS/04434/2013 \& POCI-01-0145-FEDER-007672, PTDC/FIS-AST/1526/2014 \& POCI-01-0145-FEDER-016886 and PTDC/FIS-AST/7073/2014 \& POCI-01-0145-FEDER-016880. P.F., S.B., N.C.S. e S.G.S. acknowledge support from FCT through Investigador FCT contracts nr. IF/01037/2013CP1191/CT0001, IF/01312/2014/CP1215/CT0004, IF/00169/2012/CP0150/CT0002, and IF/00028/2014/CP1215/CT0002. V.A. and E.D.M. acknowledge support from FCT through Investigador FCT contracts nr. IF/00650/2015/CP1273/CT0001, IF/00849/2015/CP1273/CT0003;and by the fellowship SFRH/BPD/70574/2010, SFRH/BPD/76606/2011 funded by FCT and POPH/FSE (EC). PF further acknowledges support from Funda\c{c}\~ao para a Ci\^encia e a Tecnologia (FCT) in the form of an exploratory project of reference IF/01037/2013CP1191/CT0001. ACSF is supported by grant 234989/2014-9 from CNPq (Brazil). J.P.F. acknowledges support from FCT through the grant reference SFRH/BD/93848/2013.
\end{acknowledgements}

%----------------------------------------------------------------------------------------
%	Bibliography 
%----------------------------------------------------------------------------------------
\bibliographystyle{aa}
\bibliography{santos_bibliography}

%----------------------------------------------------------------------------------
%	Appendices
%----------------------------------------------------------------------------------
%\appendix

%\section{Flux received by the ring}  \label{app:flux}

% Here, we detail the computation of the flux $F_r(\phi)$ received by the

\end{document}